# Analysis of the shape of x-ray diffraction peaks originating from the hexatic phase of liquid crystal films

I. A. Zaluzhnyy,[1,2*] R. P. Kurta,[3†] A. P. Menushenkov,[2‡] B. I. Ostrovskii,[4,5§] and I. A. Vartanyants[1,2¶]

[1]Deutsches Elektronen-Synchrotron DESY, Notkestrasse 85, D-22607 Hamburg, Germany

[2]National Research Nuclear University MEPhI (Moscow Engineering Physics Institute),
Kashirskoe shosse 31, 115409 Moscow, Russia

[3]European XFEL GmbH, Holzkoppel 4, D-22869 Schenefeld, Germany

[4]FSRC "Crystallography and Photonics", Russian Academy of Sciences, Leninskii prospect 59, 119333 Moscow, Russia

[5]Landau Institute for Theoretical Physics, Russian Academy of Sciences,
prospect akademika Semenova 1-A, 142432 Chernogolovka, Russia

(Dated: September 15, 2016)

## Abstract

X-ray diffraction studies of the bond-orientational order in the hexatic-B phase of 75OBC and 3(10)OBC compounds are presented. The temperature evolution of an angular profile of a single diffraction peak is analyzed. Close to the hexatic-B–smectic-A transition these profiles can be approximated by the Gaussian function. At lower temperatures in the hexatic-B phase the profiles are better fitted by the Voigt function. Theoretical analysis of the width of diffraction peaks in three-dimentional (3D) hexatics is performed on the basis of the effective Hamiltonian introduced by Aharony and Kardar [1]. Theoretical estimations are in good agreement with the results of x-ray experiments.

[*]Corresponding author: ivan.zaluzhnyy@desy.de

[†]ruslan.kurta@xfel.eu

[‡]apmenushenkov@mephi.ru

[§]ostrenator@gmail.com

[¶]ivan.vartaniants@desy.de

I. INTRODUCTION

The hexatic phase is an intermediate phase that appears in the process of two-dimensional (2D) crystal melting and was first introduced by Halperin and Nelson [2, 3]. The 2D hexatic phase is characterized by quasi-long-range bond-orientational (BO) order and short-range (liquid-like) positional order. It was predicted theoretically [2, 3] that the specific mechanism of the appearance of the hexatic phase consists in dissociation of the thermally exited dislocation pairs in a 2D crystal. This transforms the positional order from a quasi-long-range to a short-range. The subsequent dissociation of a single dislocation into two uncoupled disclinations leads to the disturbance of the BO order and makes it short-range. The hexatic phase was experimentally observed in a number of 2D systems [4-7], including some thermotropic and lyotropic liquid crystals (LCs) [8, 9]. Surprisingly the hexatic phase was observed not only in 2D LCs, but also in multilayer smectic LCs [10], where the defect-mediated mechanism does not work due to high energy cost of defects formation in 3D. This 3D hexatic phase was called a stacked hexatic phase [11], since it can be considered as a stack of parallel molecular layers possessing the hexatic ordering. In each layer the elongated LC molecules are oriented perpendicular to the layer plane (hexatic-B phase), and layers are coupled to each other. This means that the orientation of intermolecular bonds is the same in all layers, while the shear modulus between the layers is zero, so there are no positional correlations between different layers [10].

The BO order in the hexatic phase is characterized by the local order field $\psi(\mathbf{r}) \propto |\psi(\mathbf{r})|\exp[i6\theta(\mathbf{r})]$, where $|\psi(\mathbf{r})|$ is a magnitude of the two-component BO order parameter and $\theta(\mathbf{r})$ is its phase that corresponds to an angle between the intermolecular bonds and some reference axis [2]. Conventional way to obtain information about the BO order in the hexatic phase is to consider a set of the BO order parameters $C_{6m} = Re\langle\psi^m\rangle$ [12, 13]. The hexatic-B phase is characterized by non-zero values of the BO order parameters, while in absence of the BO order $C_{6m} = 0$ for all $m$. For that reason the values of the BO order parameters and the functional relation between them are important to characterize the structure

of the hexatic-B phase. The multicritical scaling theory [12, 13] predicts, that in the hexatic phase the BO order parameters obey a scaling law, $C_{6m} = C_6^{\sigma_m}$, where $\sigma_m = m + \lambda m(m-1)$. As it follows from this theory he value of the parameter λ is equal to unity for the 2D hexatic-B phase. That gives $\sigma_m = m^2$ and implies the Gaussian shape of the diffraction peaks in the direction along the ark [14, 15]. In the 3D hexatic-B phase parameter λ≈0.3, which corresponds to a more complex shape of the diffraction peak in the azimuthal direction [12, 13].

A convenient way to observe the hexatic order is to perform an x-ray scattering experiment. A diffraction pattern from a hexatic film consists of six diffraction peaks that correspond to sixfold rotational symmetry of the hexatic phase. The relative orientation between the LC molecules in the hexatic phase and the corresponding diffraction pattern are schematically shown in Figs. 1(a-b). The broadening of the diffraction peaks is caused by the presence of fluctuations of the phase $\theta(r)$, so the BO order parameters can be determined by fitting the peak profiles with a Fourier series [12, 13]. There is another method called an x-ray cross-correlation analysis (XCCA), that is based on the Fourier decomposition of the angular cross-correlation functions [16-18]. This method allows one to obtain the values of the BO order parameters directly from the measured x-ray diffraction patterns without any fitting procedure and with high precision, which is complicated to achieve by other methods [19-20]. In this work we consider a complementary approach, which is based on the analysis of the angular shape of diffraction peaks. Such an analysis of the peak profiles provides valuable information about the structure and BO order in the hexatic phase. Below we derive analytical expressions for the width of diffraction peaks in 2D and 3D hexatics and compare them with experimental observations. We also report on the x-ray diffraction study of the shape of individual peak in thick free-standing films.

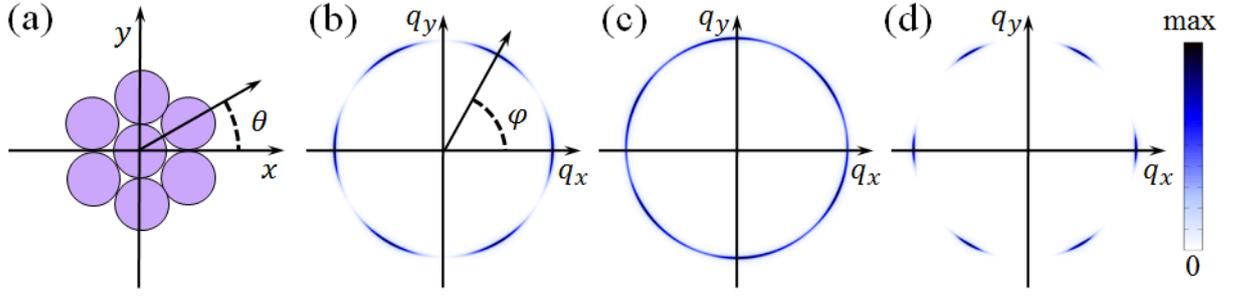

Fig. 1. (a) Top view of the LC molecules in the hexatic phase and definition of the angle $\theta$. (b-d) Schematic illustration of the diffraction patterns from a single hexatic layer (b), stack of $N = 3$ uncoupled hexatic layers (c) and $N \gg 1$ coupled hexatic layers (d).

## II. THEORETICAL ANALYSIS OF MEAN SQUARE FLUCTUATIONS OF THE BOND-ORIENTAIONAL ORDER PARAMETER

To analyze the fluctuations of the phase $\theta(r) \equiv \theta$ and derive the expression for $\langle \delta\theta^2 \rangle$ we consider a stack of $N$ parallel hexatic layers on the basis of the effective Hamiltonian introduced by Aharony and Kardar [1]

$$\mathcal{H} = \int d^2r \left[ \frac{K_A}{2} \sum_{i=1}^{N} (\nabla \theta_i)^2 + \frac{J}{2} \sum_{i=1}^{N-1} (\theta_{i+1} - \theta_i)^2 \right]. \quad (1)$$

Here $\theta_i$ corresponds to the phase of the BO order in the $i^{\text{th}}$ layers, $\nabla$ is a gradient in the plane of the layer, the Frank constant $K_A$ characterizes the effective stiffness of the BO field within the molecular layer, and $J$ is a constant describing the BO coupling between two layers. We assume, that the temperature is far away from the smectic-A-hexatic-B phase transition, so we can neglect fluctuations of the magnitude $|\psi(r)|$ of the BO order field and consider only fluctuations of the phase $\theta(r)$.

In the case of a single layer, $N = 1$, only one term remains in the sum and the Hamiltonian (1) reduces to the form corresponding to the so-called XY-model [21]. In this case the mean square of the fluctuations of the angle $\theta(\mathbf{r})$ can be expressed by an integral

$$\langle \delta\theta^2 \rangle = \frac{k_B T}{2\pi K_A} \int \frac{dq_\perp}{q_\perp} = \frac{k_B T}{2\pi K_A} \ln\frac{W}{a}, \quad (2)$$

which is logarithmically divergent [22]. Here $q_\perp$ is the component of the wavevector transfer that lies in the plane of a molecular layer, $W$ is the sample size in the plane of the layer and $a$ is an in-plane intermolecular distance. Herewith, the average value of $\psi(\mathbf{r})$ in the infinite 2D hexatic film is equal to zero. It means that the x-ray diffraction pattern from such a sample should consist of an isotropic scattering ring. However, finite-size effects modify this pattern, so that a sixfold modulation is observed in one or two layer thick monodomain hexatic-B films [14, 23]. Since thermodynamic stability of the hexatic phase requires $K_A/k_B T \geq 72/\pi$ [2, 3], one can estimate the value of $\langle \delta\theta^2 \rangle^{1/2}$ to be as large as 16° (for $W \approx 50~\mu m$, $a \approx 0.5$ nm). This causes substantial azimuthal broadening of the diffraction peaks in the 2D hexatic phase. A schematic illustration of the diffraction pattern for the case $N = 1$ is shown in Fig. 1(b). The estimate of 16° is in a good agreement with data from electron diffraction experiments on two layer thick LC films, in which the half width at half maximum (HWHM) of the hexatic peaks varies from 8° to 12° [14, 23].

Now we will consider several hexatic layers, which do not interact with each other, i.e. $N > 1$ and $J=0$. In this case the Hamiltonian (1) transforms into a sum of independent terms, each of them corresponds to a single molecular layer. In this case the mean square fluctuations $\langle \delta\theta_i^2 \rangle$ in each layer are described by the integral (2) and depend on the lateral domain size $W$, which can be different from each individual layer. The values of the mean phase $\langle \theta_i \rangle$ in this case are independent for each layer and the sixfold diffraction pattern from each layer is randomly oriented in the scattering plane. As a result the diffraction pattern from a stack of non-interacting layers will look like uniform ring of scattering. This situation is schematically illustrated in Fig. 1(c), where a diffraction pattern from $N = 3$ uncoupled hexatic layers is shown.

In the case of thick LC film consisting of many interacting layers, i.e. $N \gg 1$ and $J > 0$, one can approximate $(\theta_{i+1} - \theta_i) \approx d(\partial \theta_i/\partial z)$ and replace the sum over the layers by an integral, $\Sigma \to \int dz/d$, where the z-axis is perpendicular to the layers and $d$ is an interlayer spacing. After these transformations the Hamiltomian (1) can be written in a continuous form

$$\mathcal{H} = \int d^2r \int dz \left[ \frac{K_A}{2d} (\nabla \theta)^2 + \frac{Jd}{2} \left( \frac{\partial \theta}{\partial z} \right)^2 \right]. \quad (3)$$

We should note here that the assumptions that lead to Eq. (3) are valid if $\theta(\mathbf{r}, z)$ slowly varies on the scale of the in-plane molecular separation $a$ and interlayer distance $d$, which corresponds to long-wave fluctuations of the BO order. This type of fluctuations is dominant at low temperatures far away from the smectic-A-hexatic-B phase transition. One can show that for the Hamiltonian given by Eq. (3) the mean square amplitude $\langle \delta \theta^2 \rangle$ can be expressed by an integral in cylindrical coordinates [21]

$$\langle \delta \theta^2 \rangle = \frac{k_B T d}{(2\pi)^2} \int \frac{q_\perp dq_\perp dq_z}{K_A q_\perp^2 + J d^2 q_z^2}, \quad (4)$$

where $\pi/W \leq q_\perp \leq \pi/a$, $2\pi/L \leq q_z \leq 2\pi/d$, and $L = Nd$ is the thickness of the film. Evaluating this integral in the limit $W \gg a$ and $L \gg d$ one obtains

$$\langle \delta \theta^2 \rangle = \frac{k_B T}{4\pi K_A} \left[ \ln \left( 1 + \frac{1}{4x} \right) + \frac{1}{\sqrt{x}} \operatorname{arctg}(2\sqrt{x}) \right], \quad x = \frac{Ja^2}{K_A}. \quad (5)$$

Due to presence of the interaction between the layers in expression (3) the mean square fluctuations $\langle \delta \theta^2 \rangle$ are not divergent and approach some finite average value. For a large value of $J$ the mean square fluctuations tend to zero as $k_B T/8a\sqrt{K_A J}$. For $J \to 0$ the mean square fluctuations of the angle θ are still logarithmically divergent as it should be for a stack of 2D uncoupled hexatic layers. It is interesting to note, that the coupling parameter $J$ is present in Eq.

(5) only as a dimensionless parameter $x = Ja^2/K_A$, which characterizes anisotropy of the media. Although the final equation for the mean square fluctuations $\langle\delta\theta^2\rangle$ does not depend directly on the interlayer distance $d$, the value of the coupling parameter $J$ depends on this distance.

In order to estimate the width of the diffraction peak we have to define some value of the coupling parameter $J$. Assuming that contributions to the energy from two terms in the Hamiltonian (1) are equal, one can estimate the value of $J$ as $J \sim K_A/a^2$ [1]. In this case for $K_A/k_BT = 72/\pi$ and $a = 0.5$ nm one obtains from Eq. (5) that $\langle\delta\theta^2\rangle^{1/2} \approx 4°$. A schematic illustration of the diffraction pattern in the case of $N \gg 1$ and $J > 0$ is shown in Fig. 1(d). The diffraction peaks are more narrow in the azimuthal direction in comparison with the 2D case. It should be noted that the above estimates have to be considered with a certain caution. We have used in our calculations the minimum value of the Frank elastic constant $K_A/k_BT \geq 72/\pi$. Actually, for concrete LC compounds the values of the elastic constant $K_A$ and the interlayer coupling parameter $J$ can be larger than assumed above. This can lead to smaller values of the mean square amplitude $\langle\delta\theta^2\rangle$.

## III. X-RAY DIFFRACTION EXPERIMENT

We used 75OBC and 3(10)OBC LC compounds that belong to *n*-alkilyl-4'-*n*-alkoxybiphenyl-4-carboxylate series and exhibit similar phase transition sequences: Sm-A(63.8° C)-Hex-B(below 59° C)-Cr-E (75OBC) and Sm-A(66.3° C)-Hex-B(below 54° C)-Cr-E (3(10)OBC) [24, 25]. The temperature range of the hexatic-B phase existence is more than two times larger for 3(10)OBC compound. To compare the results obtained for different compounds we will use a relative temperature $t = T - T_C$, counted from the smectic-A-hexatic-B phase transition temperature $T_C$. Freely suspended LC films of the thickness within the range from 5 to 8 μm were prepared in the smectic-A phase and cooled down to hexatic-B phase and then finally to the crystal-E phase. At each temperature we collected the diffraction signal from different spatial positions on the sample, which has been used later to obtain an averaged diffraction patterns for better statistics.

The experiment was performed at Coherence Beamline P10 of PETRA III synchrotron source (DESY, Hamburg) at photon energy E=13 keV (see for experimental details [19, 20]). The LC film was oriented perpendicular to the incident x-ray beam and the diffraction patterns were measured with a Pilatus 1M detector. The pixel size of the detector (172x172 $\mu m^2$) corresponds to angular resolution of $\Delta\varphi = 0.2°$ per pixel. The first-order diffraction peaks occur at approximately $q_0 = 1.43$ nm$^{-1}$, which corresponds to the in-plane intermolecular spacing of $a \approx 0.5$ nm [26].

## IV. EXPERIMENTAL STUDIES OF THE SHAPE AND WIDTH OF THE ANGULAR PROFILES

It is convenient to use the polar coordinates $(q, \varphi)$ in the detector plane to study diffraction patterns from the hexatic phase (see Fig. 1(a)). Below we will analyze the angular dependence of the scattered intensity I(q$_0$,φ) at a fixed value of the wavevector transfer $q = q_0$, which is denoted as $I(\varphi)$. The radial cross section through the diffraction peak can be described by the Lorentzian function, that corresponds to the short-range in-plane positional order (see for details [19, 20, 27]).

The angular dependence $I(\varphi)$ was obtained by using the following procedure. First, the position of the center of the diffraction pattern and the value of scattering vector $q_0$ were adjusted in such a way, that the circle of a radius $q_0$ passes through all the maxima of the diffraction peaks (position of the center of the diffraction pattern does not change with the temperature, while the value of $q_0$ slightly increases on cooling [20]). Then the angular dependence $I(\varphi)$ was determined from the experimentally measured diffraction patterns. Examples of the angular dependence $I(\varphi)$ for 3(10)OBC compound measured at a single position on the sample are shown in Fig. 2. At the relative temperature $t = -0.8$ °C close to the smectic-A-hexatic-B phase transition (Fig. 2(a)) the diffraction peaks are broad and noisy and the maximum intensity is only about 25 counts per pixel. At the lower temperature $t = -9.3$ °C (Fig. 2(b)) the diffraction peaks become much sharper and the signal becomes stronger and the maximum intensity reaches about 250 counts per pixel. The similar temperature behavior was observed for 75OBC compound.

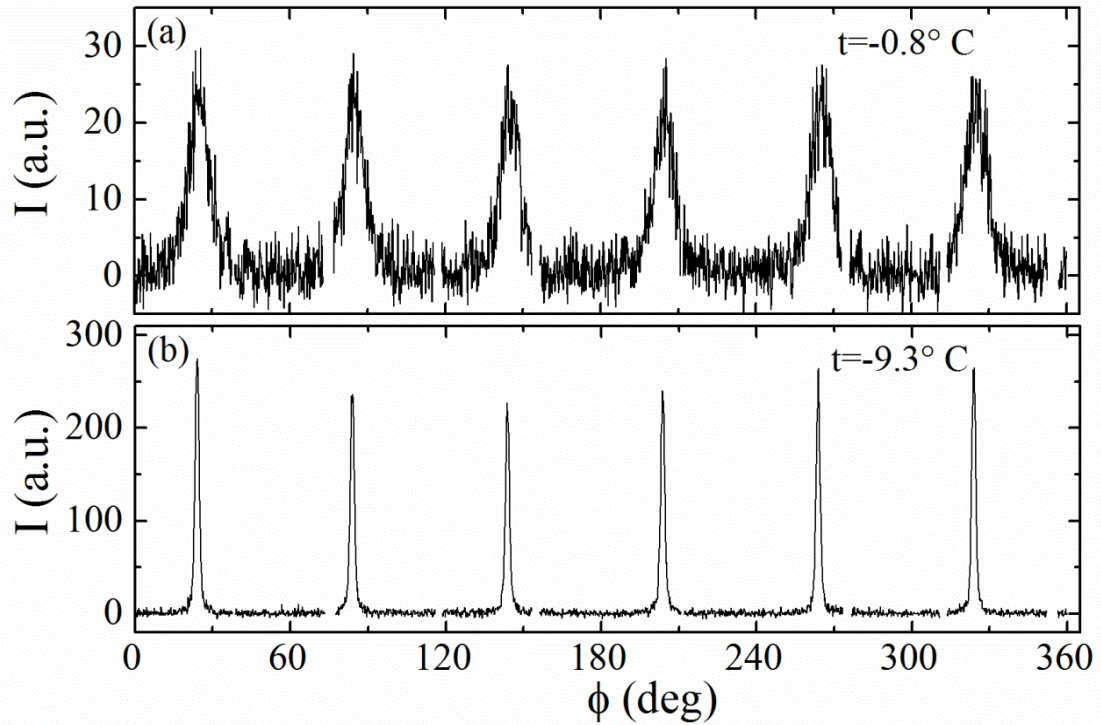

FIG. 2. Angular dependence of the scattered intensity in 3(10)OBC film measured at one position on the sample at the relative temperature $t = -0.8$ °C (a) and $t = -9.3$ °C (b) in the hexatic phase.

One can notice that six diffraction peaks at the relative temperature $t = -9.3°$ C have slightly different magnitude, which we attribute to misalignment of the experimental setup. At the same time all peaks have the same shape, so one can study the BO order by analyzing a profile of any of the diffraction peaks. In Figs. 3 and 4 averaged peak profiles are shown at different temperatures for 3(10)OBC and 75OBC compounds respectively. The angular profiles close to the phase transition temperature can be well approximated by a Gaussian function. Gaussian profile is predicted for 2D hexatics by the multicritical scaling theory [14, 15, 28]. However, the scaling law for 3D hexatics, $C_{6m} = C_6^{\sigma_m}$, where $\sigma_m = m + \lambda m(m-1)$ and $\lambda \approx 0.3$, implies a more complex angular shape of the diffraction peak. The angular peak profile in thick hexatic films can be well described by the Voigt function $V(\varphi; \sigma, \gamma) = G * L$, that is a convolution of the Gaussian function $G(\varphi) = \exp\left(-\frac{\varphi^2}{2\sigma^2}\right)/\sqrt{2\pi\sigma^2}$ and the Lorentzian function $L(\varphi) = \gamma/\pi(\gamma^2 + \varphi^2)$, where the parameters σ and γ define the width of the Gaussian and Lorentzian functions [29]. At high temperatures close to the smectic-A-hexatic-B phase transition the Gaussian function gives rather good fit of the experimental data, however at lower temperatures deep in the hexatic phase the Voigt function provides much better fit of the experimental profiles, especially on the tails of the peaks (see Figs 3 and 4).

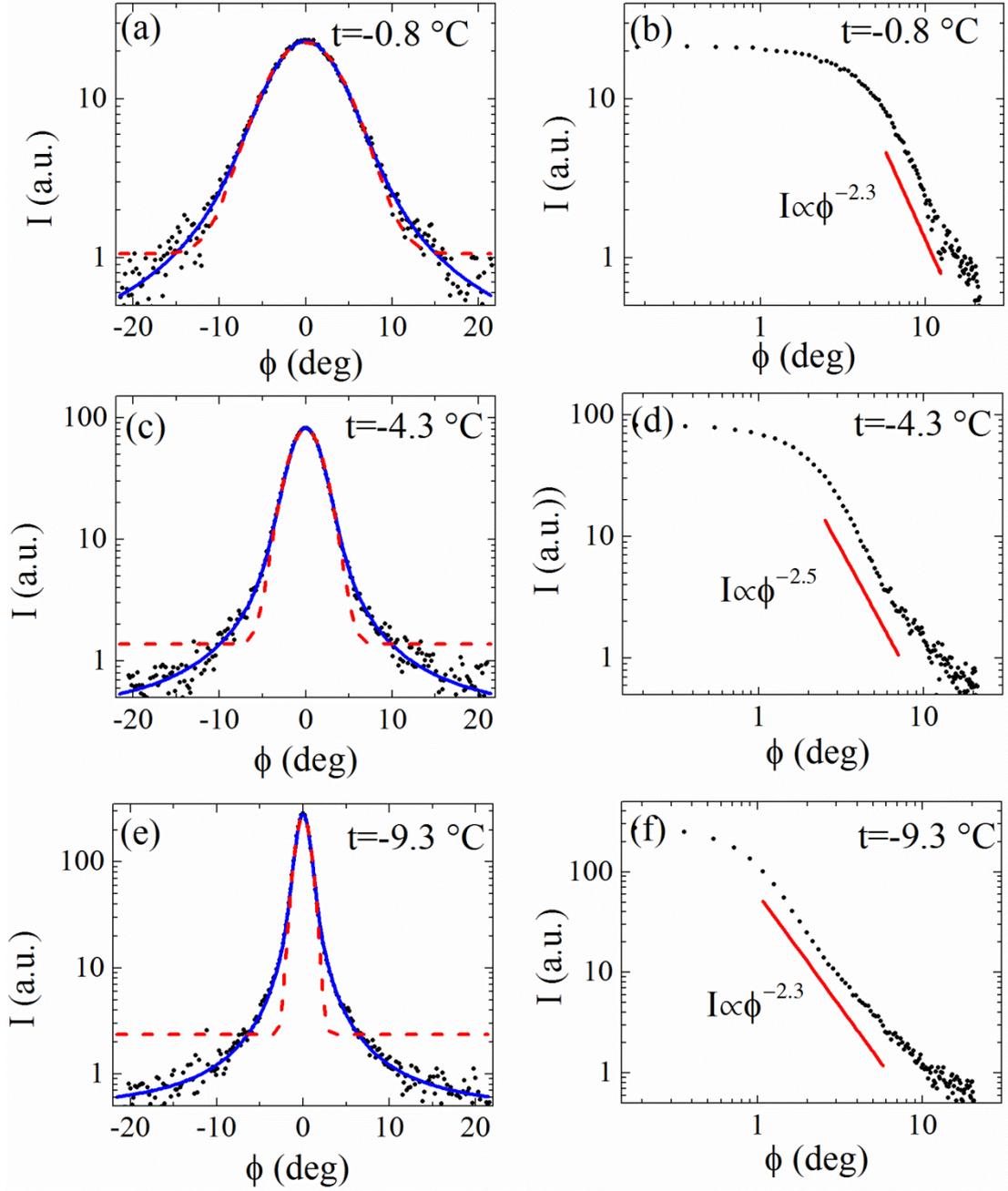

FIG. 3. Azimuthal profile of a diffraction peak in the hexatic phase for 3(10)OBC sample averaged over 100 diffraction patterns at $t = -0.8\,°C$ (a,b), $t = -4.3\,°C$ (c,d) and $t = -9.3\,°C$ (e,f) below the smectic-A-hexatic-B phase transition. Experimental data are shown with black points. In the linear-log figures (a,c,e) fitting of the experimental data with Gaussian profile and Voigt profile are shown with dash red and solid blue lines, respectively. In the log-log figures (b,d,f) red lines show the power decay of intensity.

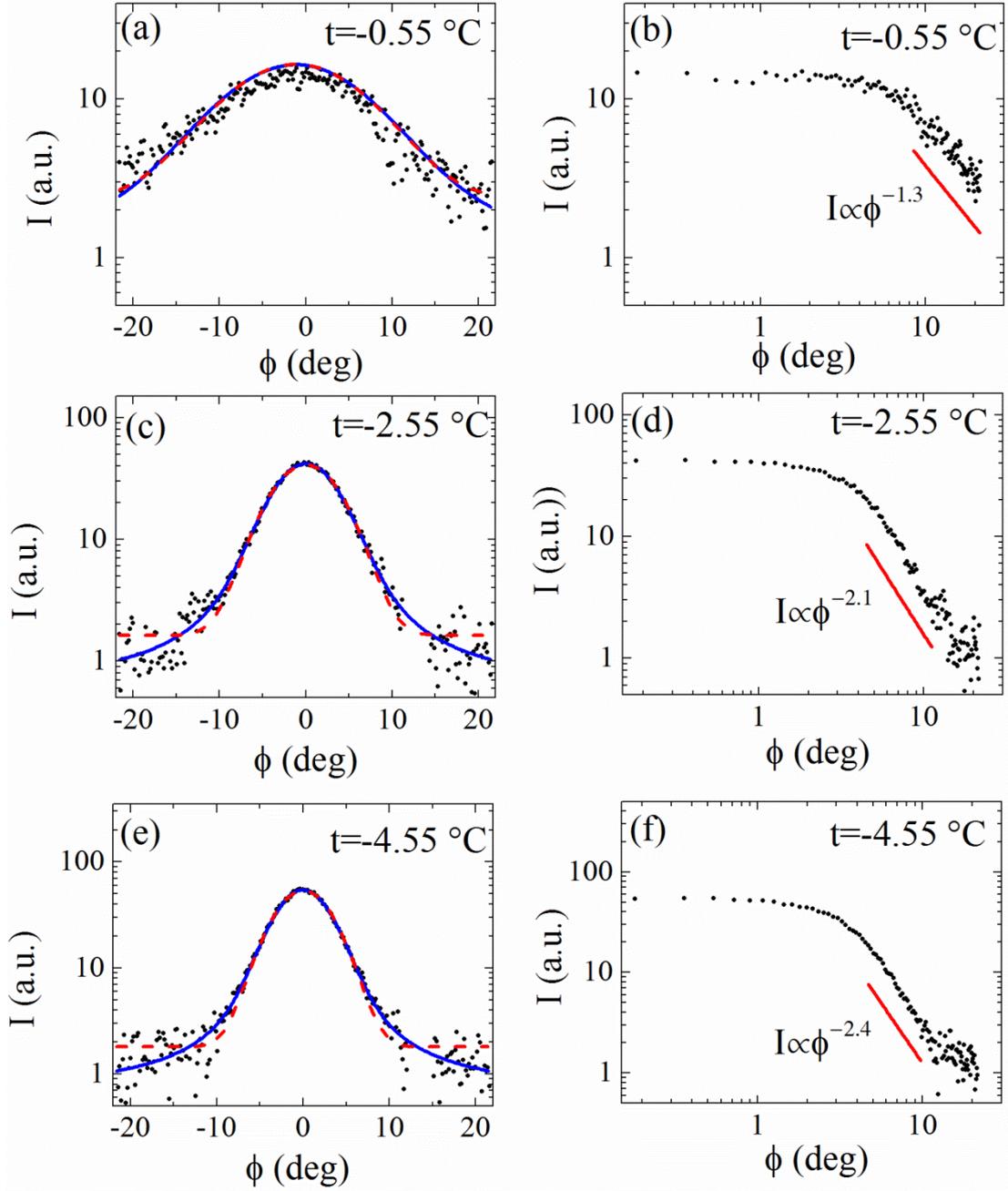

FIG. 4. Azimuthal profile of a diffraction peak in the hexatic phase for 75OBC sample averaged over 25 diffraction patterns at $t = -0.55$ °C (a,b), $t = -2.55$ °C (c,d) and $t = -4.55$ °C (e,f) below the smectic-A–hexatic-B phase transition. Experimental data are shown with black points. In the linear-log figures (a,c,e) fitting of the experimental data with Gaussian profile and Voigt profile are shown with dash red and solid blue lines, respectively. In the log-log figures (b,d,f) red lines show the power decay of intensity.

To characterize the relative contributions of Gaussian and Lorentzian functions to the profile of the diffraction peak fitted by the Voigt function we consider a dimensionless parameter $\rho = \Gamma_L/(\Gamma_L + \Gamma_G)$, where $\Gamma_L = 2\gamma$ and $\Gamma_G = 2\sqrt{2\ln 2}\,\sigma$ are a full width of half maximum (FWHM) of Lorentzian and Gaussian functions, respectively [29]. In the case of $\Gamma_L = 0$ the Voigt function transforms to a pure Gaussian function, while in the case of $\Gamma_G = 0$ it reduces into a pure Lorentzian function. Parameter $\rho$ varies from zero, that corresponds to the Gaussian function, to unity, that corresponds to the Lorentzian function. The temperature dependence of the parameter $\rho$ for both 3(10)OBC and 75OBC compounds is shown in Fig. 5. The value of $\rho$ lies in the range from 0.25 to 0.4 for both compounds, that corresponds to predominant Gaussian contribution, and slightly increases upon cooling.

An important information about the structure of the hexatic-B phase can be obtained from the analysis of the tails of the diffraction peaks. The decay of the intensity on the tails can be well described by power the law $I \propto \varphi^{-n}$, where $n \approx 1.0 - 2.5$ at all temperatures (see Figs. 3(b,d,f) and 4(b,d,f)). If broadening of the diffraction peaks is caused by the presence of dislocations, as one may expect for 2D hexatics, the intensity on the tails of the diffraction peaks will decrease as a power law with $n \approx 3 - 4$ [30], which is clearly not the case in our experiment.

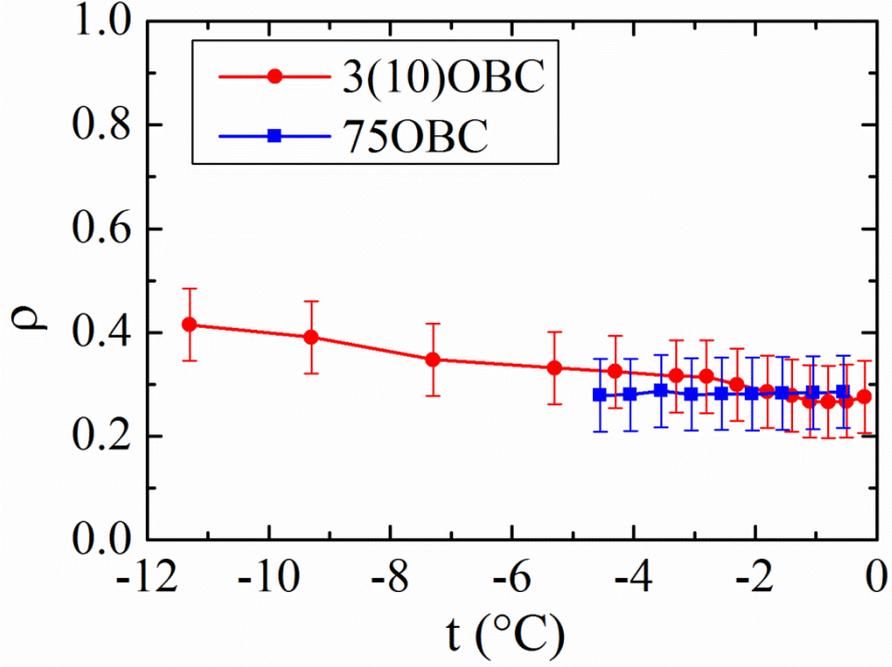

Fig. 5. Temperature dependence of the parameter ρ obtained from the fitting of the averaged diffraction peak with the Voigt function.

Now we will consider the mean square fluctuations $\langle \delta\theta^2 \rangle$ by analyzing the width of the diffraction peaks. Regardless of the shape of the diffraction peaks one can consider a variance

$$\langle \delta\varphi^2 \rangle = \frac{\int I(\varphi)(\varphi - \varphi_0)^2 d\varphi}{\int I(\varphi) d\varphi}, \qquad (6)$$

where $I(\varphi)$ is angular dependence of the scattered intensity of a single diffraction peak and $\varphi_0$ is the angular position of the center of this peak. The standard deviation $\langle \delta\varphi^2 \rangle^{1/2}$ is directly related to the width of the diffraction peak. For example, in the case of a Gaussian profile FWHM $\Gamma_G$ is proportional to the standard deviation, $\Gamma_G = 2\sqrt{2 \ln 2}\, \langle \delta\varphi^2 \rangle^{1/2}$. Fluctuations of the BO order parameter causes corresponding broadening of the diffraction peaks in the hexatic-B phase, so we can assume that $\langle \delta\varphi^2 \rangle = \langle \delta\theta^2 \rangle$.

The temperature dependence of the standard deviation $\langle \delta\varphi^2 \rangle^{1/2}$ determined from the analysis of the diffraction peaks for 3(10)OBC and 75OBC compounds is shown in Fig. 6. We observed the same trend for both compounds, the peaks are wider in the region close to phase transition and become narrow at lower temperatures. At the same time we noticed that the peaks for 75OBC compound are wider, which means that fluctuations of the BO order are larger in this system. At lower temperatures deep in the hexatic-B phase the value of mean square fluctuations does not depend strongly on the temperature and nearly saturates. The values of $\langle\delta\varphi^2\rangle^{1/2}$ are about 5° for 75OBC compound and about 3° for 3(10)OBC compound at the lowest temperatures of the hexatic-B phase existence. These values are in a good agreement with the theoretically estimated value of 4° presented in section II above. From this observation we can conclude, that our assumption for the value of the coupling parameter $J \sim K_A/a^2$ seems to be valid in our experiment. The lower value $\langle\delta\varphi^2\rangle^{1/2}$ for 3(10)OBC compound indicates that for this compounds intermolecular bonds are more rigid and the Frank constant $K_A$ is larger, as compared to 75OBC compound.

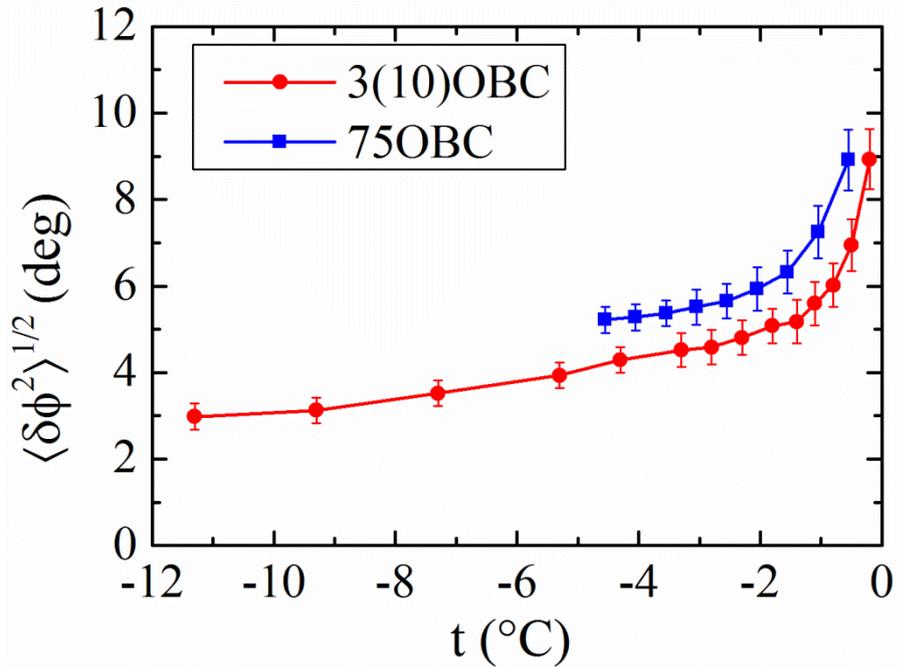

FIG 6. Temperature dependence of the standard deviation $\langle\delta\varphi^2\rangle^{1/2}$ for the averaged azimuthal profile of a diffraction peak in the hexatic-B phase.

IV. CONCLUSIONS

In summary, we investigated the diffraction peaks width and the features of the BO order in thick LC hexatic-B films of 3(10)OBC and 75OBC compounds. It is shown that angular profiles of the diffraction peaks are well described by the Voigt function in a whole temperature range of the hexatic-B phase existence. On the tails of the diffraction peaks the intensity decays as $I(\varphi) \propto \varphi^{-n}$, with the power $n \approx 1.0 - 2.5$. The temperature dependence of the width of the diffraction peaks was determined from the experimental x-ray diffraction data. Theoretical analysis of the width of the diffraction peaks in 3D hexatics was performed. The results of theoretical analysis are in a good agreement with the experimental data.

ACKNOLEDGEMENTS

We acknowledge the discussions and support of this project by E. Weckert. We are grateful to the stuff of the Coherence Beamline P10 at PETRA III, especially to M. Sprung and A. Zozulya, for the help during experiment. We thank E. I. Kats, V. V. Lebedev, A. R. Muratov, E. S. Pikina and V. M. Kaganer for fruitful discussions. This work was partially supported by the Virtual Institute VH-VI-403 of the Helmholtz Association. The work of I.A.Z. and B.I.O. was partially supported by the Russian Science Foundation (Grant No.14-12-00475).